\documentclass[preprint,aps,superscriptaddress,nofootinbib]{revtex4-2}
\usepackage{graphicx}
\usepackage{epsfig}
\usepackage{bm}
\usepackage{amssymb}
\usepackage{wrapfig}
\usepackage{CJKutf8, here}

\usepackage[usenames, dvipsnames]{color}

\newcommand{\cA}{{\cal A}}

\newcommand{\el}{{\cal L}}
\newcommand{\cM}{{\cal M}}

\newcommand{\cT}{{\cal T}}

\newcommand{\cW}{{\cal W}}

\newcommand{\hpsi}{\mbox{${\hat \psi}$}}
\newcommand{\hpsibar}{\mbox{$\bar{\hat \psi}$}}

\newcommand{\tldF}{\mbox{${\tilde F}$}}

\newcommand{\epsi}{\mbox{$\varepsilon$}}

\newcommand{\vq}{\mbox{$\bm{q}$}}
\newcommand{\vbr}{\mbox{$\bm{r}$}}

\newcommand{\vB}{\mbox{$\bm{B}$}}

\begin{document}

\title{High Energy Particle Production from Proton Synchrotron Radiation 
in Strong Magnetic Fields in Relativistic Quantum Field Theory}
\author{Tomoyuki~Maruyama}
\email{maruyama.tomoyuki@nihon-u.ac.jp}
\affiliation{College of Bioresource Sciences,
Nihon University,
Fujisawa 252-8510, Japan}
\affiliation{ Department of Physics, 
Tokyo Metropolitan university, Hachioji, Tokyo 181-8588, Japan}
\author{A.~Baha~Balantekin}
\affiliation{Department of Physics, University of Wisconsin, Madison,
WI 53706, USA}
\author{Myung-Ki~Cheoun}
\affiliation{Department of Physics and Origin of Matter and Evolution of Galaxies (OMEG) Institute, Soongsil University, Seoul, 156-743, Korea}
\affiliation{National Astronomical Observatory of Japan, 2-21-1 Osawa, 
Mitaka, Tokyo 181-8588, Japan}
\author{Akira~Dohi}
\affiliation{RIKEN Pioneering Research Institute (PRI), 2-1 Hirosawa, Wako, Saitama 351-0198, Japan}
\affiliation{RIKEN Center for Interdisciplinary Theoretical \& Mathematical Sciences (iTHEMS), RIKEN 2-1 Hirosawa, Wako, Saitama 351-0198, Japan}
\author{Ryo~Higuchi}
\affiliation{RIKEN Pioneering Research Institute (PRI), 2-1 Hirosawa, Wako, Saitama 351-0198, Japan}
\author{Toshitaka~Kajino}
\affiliation{School of Physics, Peng Huanwu Collaborative Center for Research and Education, and International Research Center for Big-Bang Cosmology and Element Genesis, Beihang University, Beijing 100191, China}
\affiliation{Graduate School of Science, The University of Tokyo, 7-3-1 Hongo, Bunkyo-ku, Tokyo 113-033, Japan}
\affiliation{ Division of Science, National Astronomical Observatory of Japan, 2-21-1 Osawa, Mitaka, Tokyo 181-8588, Japan}
\author{Grant J. Mathews}
\affiliation{Center of Astrophysics, Department of Physics and Astronomy,
University of Notre Dame, Notre Dame, IN 46556, USA}

\date{\today}


\begin{abstract}
We investigate photon, pion, and $\rho$-meson production from proton synchrotron radiation 
in the presence of strong magnetic fields. 
The proton decay widths and the luminosities of the emitted particles are calculated 
within a relativistic quantum framework that incorporates Landau quantization. 
A scaling rule is derived for the transition probability between different Landau levels.
This allows an evaluation of transitions for extremely high Landau numbers exceeding $10^{15}$. 
Furthermore, we calculate the momentum distribution of the emitted particles 
by properly including the proton recoil effect associated with particle emission. 
The results  differ significantly from conventional semiclassical approaches.
\end{abstract}

\maketitle


\newpage


\section{Introduction}
Magnetars are extremely magnetized objects, whose magnetic fields are stronger than $\sim 10^{14}\:\rm G$ \citep{ARmagnetar}. 
As the multi-messenger high-energy source, they have been in the spotlight. 
They are associated with such phenomena as strong X-ray emissions, soft $\gamma$-ray emission, and  possibly  with $\ gamma$-ray bursts (GRBs), and fast radio bursts (for a review, see \cite{2024FrASS..1188953N}). 
A fast spining magnetar is also considered as a possible source of observed ultra-high-energy cosmic rays (UHECRs) \citep{Arons03} (see also \cite{2025UHECR} for candidates of UHECR), such as the Oh-My-God (OMG) particle (3.2$\times10^{20}~{\rm eV}$) detected by the Fly's Eye Detector~\cite{Oh-My-God} and the Amaterasu particle (2.4$\times10^{20}~{\rm eV}$) detected by the Telescope Array~\cite{Amaterasu}.

Among various radiation processes, synchrotron radiation has been proposed as a source for high-energy photons 
in the GeV $-$ TeV range \cite{Gupta07,Boettcher98,Totani98,Fragile04,Asano07,Asano09}, 
and is believed to be an important emission process in various kinds of neutron stars including magnetars 
(for a review, see \cite{ARmagnetar}). 
Recently, Zhang et al. \cite{Zhang23} studied the role of proton synchrotron radiation 
in explaining very high-energy (VHE) gamma-ray emission from the exceptionally bright source GRB 221009A. 
Their model suggests that protons accelerated in the reverse shock region can emit synchrotron radiation, 
accounting for photons with energies up to $\sim 18$ TeV. 
Thus, there is motivation for continuing to study the properties of proton synchrotron radiation.

To investigate synchrotron emission in objects with strong magnetic fields, however, 
quantum-mechanical effects are not negligible. 
This is because the unique environment of magnetars makes processes 
that are forbidden in the absence of a strong magnetic field possible.  
These include $\nu\bar{\nu}$-pair synchrotron radiation~\cite{2020PhLB..80535413M} and 
the nucleon direct Urca process in low-mass magnetars~\cite{2022PhLB..82436813M}. 
In a strong magnetic field environment (typically above the Schwinger limit of 
$B\gtrsim4.4\times10^{13}~{\rm G}$), 
a classical calculation \citep{TK99} cannot be applied due to Landau quantization. 
This is because despite the continuous parameter of the magnetic field $B$ in the classical calculation, 
the quantum-mechanical treatment involves discrete values of the magnetic field $B$. 
The synchrotron emission can occur through any quanta that may couple to an accelerated particle.  
Since protons strongly couple to meson fields, a high-energy proton can also radiate pions and other mesons, 
as well as photons. 
The meson-nucleon couplings are about 100 times larger than the photon-nucleon coupling, 
Hence, the meson production process is expected to exceed photon synchrotron emission 
in the high-energy regime. 
For example, $\pi^0$  emission from synchrotron radiation is discussed in Refs. \cite{Ginzburg65a,Ginzburg65b,Zharkov65,TK99,BDK95}, and $\rho^0$ production is further discussed in Ref.~\cite{Kajino14}. However, these calculations were performed in a semi-classical approximation and invoked an approximate quantum-mechanical treatment of the proton transitions among the Landau levels associated 
with the strong magnetic field.
In our previous work \cite{P2Pi-1}, we exploited the Green's function method for the propagation
of protons in a strong magnetic field.
We studied the pion production  from proton synchrotron emission in a relativistic quantum field-theory approach,
whereby the pion is produced from the transition of a proton between two Landau levels.
We then deduced the energy and angular distribution of emitted pions
which had not been deduced in previous semi-classical approaches.
However, to keep the Landau level number to a value that allows actual numerical calculations, we had to set the magnetic field to about $10^{18}$~G and the incident proton energy to less than 10 GeV.

In our next work \cite{P2PiHe} we found that the proton pionic decay width satisfies a scaling law
when using the PV coupling for the nucleon-pion interaction.
Then, we can calculate the proton decay width and momentum distribution, including the angular dependence
up to a few hundred GeV.
However, the obtained decay width increases monotonically  
as the incident  proton energy becomes larger,
and the pion luminosity is very large in the very high energy region because of the PV coupling. 

On the other hand, the proton decay widths do not satisfy the scaling law for the photon
and vector-meson emissions.  
In addition, the decay width for pion emission with the PS coupling does not satisfy it either.
Therefore, we could not develop our method further at that time.

However, in this paper we report a more general scaling rule and a new approach to calculate the proton decays
to any particle within the relativistic quantum framework. 
Here, we elaborate on the new approach and perform calculations for the direct photon, pion,
and $\rho$-meson emission.

Section 2 is devoted to an introduction to our theoretical formalism based upon the Landau quantization. 
Numerical results are presented 
in Section 3 along with detailed discussions. 
A summary and conclusions are presented in Section 4.   

\newpage

\section{Formalism }

In this section we briefly explain our formalism for particle production from
proton synchrotron radiation.

\subsection{Proton Field}
We assume a uniform magnetic field along the $z$-direction,
$\vB = (0,0,B)$, and take the electro-magnetic vector potential $A^{\mu}$ to be
$A = (0, 0, x B, 0)$ at the position $\vbr \equiv (x, y, z)$ .
The proton field $\hpsi(t; \vbr)$ is then obtained with the following Dirac equation:
\begin{equation}
\left[ \gamma_\mu (i \partial^\mu - e A^\mu) - M_p \right] \hpsi (t; \vbr) = 0 ,
\label{DirEq}
\end{equation}
where $M_p$ is the proton mass, and $e$ is the elementary charge.
We define $p_y$ and $p_z$ as the $y$ and $z$-components of the proton momentum,
and  the above field $\hpsi(t; \vbr)$ is  proportional to $\exp [i(p_y y + p_z z - E t)]$. 
The single particle energy, $E$, is given by
\begin{equation}
E^2 = p_z^2 + E_T^2 = p_z^2 + 2n_L eB + M_p^2 ,
\end{equation}
where $n_L$ is the proton Landau number, and $p_z$ is the $z$-component of the proton momentum.

Using the above fileds, we can write the proton Green function as
\begin{equation}
G(p_0, p_z, x, x^\prime)  =  \sum_{n_L=0} \tldF_{n_L} (\xi)  
\frac{ \rho_M (p_0,p_z, n_L) }{p_0^2 - E(n_L, p_z)^2  + i \delta}
 \tldF_{n_L} (\xi^\prime)  
\end{equation}
with
\begin{eqnarray}
&& \rho_M (p_0, p_z, n_L)  =   p_0 \gamma_0 - p_z \gamma^z + \sqrt{2 eB} n_L \gamma^2 + M_p ,
%
\\ && \tldF = {\rm diag} \left( f_{n_L}, f_{n_L-1}, f_{n_L}, f_{n_L-1} \right)
= f_{n_L}  \frac{1 + \Sigma_z }{2} + f_{n_L-1}  \frac{1 - \Sigma_z }{2}  ,
\end{eqnarray}
where $\xi (\xi^\prime) = \sqrt{eB} x (x^\prime) - p_y/\sqrt{eB}$ and $\Sigma_z=diag(1, -1, 1, -1)$.



\subsection{Proton Decay Width in a Magnetic Field}

Now, we consider the following interaction Lagrangian density for the pion ($\pi^0$)
and vector particles such as the photon ($\gamma$) and  $\rho$-meson ($\rho^0$):
\begin{equation}
 \el_\pi = g_\pi \hpsibar \gamma_5 \hpsi \phi_{\pi}  ,
 \quad
 \el_V = g_V \hpsibar \gamma_\mu \hpsi \phi_V^\mu ,
\end{equation}
where $g_\pi$ is a coupling constant between the nucleon and $\pi^0$,
$g_V$ is a coupling constant between the nucleon and a vector particle.
Here, $\hpsi$ represents the proton field operator in Eq.~(\ref{DirEq}), 
and $\phi_{\pi,V}^\alpha$ denotes the field operator of the emitted pion and vector particle, 
with $\alpha$ indicating its component. 
The momentum of the emitted particle is written as
$q=(e_q, 0, q_T, q_z)$, where  without loss of generality,  
the transverse pion momentum is assumed to be directed along the $y$-axis.

The decay width of a proton with momentum $p_z$ and Landau levels $n_i$ and $n_f$ is given by  
\begin{equation}
\frac{d^3 \Gamma_{A}}{d q^3} =  
- \frac{ g_A^2 }{2^4 \pi^2 e_q} \sum_{n_f} \frac{\delta(E_f + e_q - E_i)}{E_i E_f} \cT_{if} ,
\label{difDc}
\end{equation}
where,
\begin{eqnarray}
\cT_{if} &=&  \frac{1}{4}  \sum_{\alpha, \beta} 
G_A^{\alpha \beta}  {\rm Tr} \left\{
\rho_M (E_f, p_z - q_z, n_f) J_\alpha \rho_M (E_i, p_z, n_i) \bar{J}_\beta
\right\}  .
\label{TMat}
\end{eqnarray}
Here, $e_q$ is the energy of the emitted particle,  and
\begin{eqnarray}
J_{\alpha} &=& \int d x \, \tldF(n_f, x - q_T/2 \sqrt{eB} ) \gamma_\alpha  \tldF(n_i,  x + q_T/2 \sqrt{eB} ) ,
\nonumber \\
\bar{J} _{\beta} &=& \int d x \, \tldF(n_i, x + q_T/2 \sqrt{eB})  \gamma_\beta \tldF(n_f, x - q_T/2 \sqrt{eB}) .
\label{DefJ}
\end{eqnarray}
In addition, $G_A^{\alpha \beta}$ is a parameter depending on the interaction channel, and its details are written below.

We next introduce the definition
\begin{equation}
\cM (n_i, n_f) = \int dx f_{n_i} \left(x - \frac{q_T}{2} \right) f_{n_f} \left(x + \frac{q_T}{2} \right)  ,
\label{TrStM}
\end{equation}
and write,
\begin{eqnarray}
\cM_1 = \cM \left( n_i, n_f \right) , && \cM_2 = \cM \left( n_i - 1, n_f - 1 \right) ,
\nonumber \\
\cM_3 = \cM \left( n_i-1, n_f \right) , && \cM_4 = \cM \left( n_i, n_f - 1 \right) .
\label{TrStMC}
\end{eqnarray}

For the pion, we use $\gamma_5$ for $\gamma_\alpha$ and write
\begin{eqnarray}
&& J^5 = \left[ \cM_1 \frac{1 + \Sigma_z}{2} + \cM_2 \frac{1 - \Sigma_z}{2} \right] \gamma^5 ,
\end{eqnarray}
then the transition matrix becomes
\begin{equation}
\cT_{if} (\pi) =  \frac{\left( \cM_1^2 + \cM_2^2 \right)}{2} \left( E_i E_f - p_{iz} p_{fz} - M_p^2  \right)
- 2 \cM_1 \cM_2 eB \sqrt{n_i n_f} .
\end{equation}

For vector particles such as photons and $\rho$ mesons, we write $G_A$ as $G_A^{\mu \nu}$
($\mu,\nu = 0 - 3$) and 
\begin{equation}
G_V^{\mu \nu} = \sum_{a} \epsi^{\mu}(a) \epsi^{\nu} (a) ,
\label{VPG}
\end{equation}
where $\epsi_{\mu}$ is the polarization vector satisfying $\epsi \cdot q = 0$.

In this work we consider reactions in the UHE region.  
Hence, we can omit the longitudinal part for vector mesons as well as for the  photon.
The two transverse polarizaton vectors can be taken to be  $\epsi (1) = (0; 1, 0, 0)$ and $\epsi(2) = (0; 0, q_z, -q_T)/|\vq|$,
and the matrices $J^\mu$ and ${\bar J}^\mu$ (\ref{DefJ}) are written as
\begin{eqnarray}
J^3 =  \left[ \cM_1 \frac{1 + \Sigma_z}{2}  + \cM_2 \frac{1 - \Sigma_z}{2} \right] \gamma^3 ,
&~~&
J^2 = \left[ \cM_3 \frac{1 + \Sigma_z}{2}  + \cM_4 \frac{1 - \Sigma_z}{2} \right]  \gamma^2 ,
\nonumber \\
{\bar J}^3 = \left[ \cM_1 \frac{1 + \Sigma_z}{2}  + \cM_2 \frac{1 - \Sigma_z}{2} \right]  \gamma^3 ,
&~~&
{\bar J}^2 = \left[ \cM_4 \frac{1 + \Sigma_z}{2}  + \cM_3 \frac{1 - \Sigma_z}{2} \right] \gamma^2 .
\label{JVec}
\end{eqnarray}
{Substituting Eqs.~(\ref{VPG}) and (\ref{JVec}) into Eq.~(\ref{TMat}), we can obtain $\cT_{if}$ as
\begin{eqnarray}
\cT_{if} (V) &=&  
 \frac{1}{4} {\rm Tr} \left\{
\rho_M (E_i - e_q, 0, n_f) J^2 \rho_M (E_i, 0, n_i) \bar{J}^2
+ \rho_M (E_i-e_q, 0, n_f) J^3 \rho_M (E_i, 0, n_i) \bar{J}^3 \right\} 
\nonumber \\
& = &  \left[ \frac{ \cM_1^2 + \cM_2^2}{2} \left( 1 + \frac{q_z^2}{|\vq|^2} \right)+ \frac{\cM_3^2 + \cM_4^2 }{2} \frac{q_T^2}{|\vq|^2} \right]
\left( E_i E_f  - p_{iz} p_{fz} - M_p^2  \right)
\nonumber \\ && \qquad
- \left[ \cM_1 \cM_2\left( 1 + \frac{q_z^2}{|\vq|^2} \right)  + \cM_3 \cM_4 \frac{q_T^2}{|\vq|^2}\right] eB \sqrt{n_i n_f} .
\end{eqnarray}

Next, we discuss the treatment of the ultra-high-energy (UHE) region, $E_i \gtrsim 100$~GeV.  
In this energy range, the Landau numbers of the  incident and final protons 
are extremely large, $n_i \gtrsim 10^8$.  
Thus, we assume that the quantities $\cM_{\alpha}$ are equal and define  
$\cM = \cM_1 = \cM_2 = \cM_3 = \cM_4. $  
Then, $\mathcal{T}_{if}$ is simply given by  
\begin{equation}
\cT_{if} = d_A \cM^2 \left( E_i E_f - p_{iz} p_{fz} - 2 eB \sqrt{n_i n_f} - M_p^2 \right)
\equiv d_A \cM^2 \cA_{if} ,
\label{Tif}
\end{equation}  
where $d_A = 1$ for pions and $d_A = 2$ for vector particles.  

Here, we set $p_{iz}= 0$ and derive the decay width for this case in the following discussion.  
The decay width for $p_{iz} \neq 0$ can then be obtained by applying a Lorentz transformation in the $z$-direction.  

In the UHE region the Landau number is very large, and we cannot directly solve $\cM(n_i, n_f)$.  
To address this, we introduce the curvature parameter
\begin{equation}
\chi_p = \frac{e B E_i}{M_p^3} ,
\end{equation}
which effectively characterizes the system in this regime.  
This parameter has been widely used in semiclassical approaches~\cite{Ginzburg65a,Ginzburg65b,Zharkov65,TK99,BDK95},  
though those calculations typically emphasize the magnetic-field independence of 
either decay widths or luminosities.


In Ref.~\cite{P2PiHe}, we showed that the pion decay width exhibits a quantum scaling behavior 
governed by $\chi_p$.  
However, this scaling is not generally valid for all particle emission processes.  
To go beyond these limitations, we propose a more general framework.

We define
\begin{equation}
 \cW_{if} = \sqrt{\frac{n_i}{eB}} \int \frac{d q_z}{2 \pi} \left[ \cM(n_i, n_f) \right]^2 ,
\label{Wif}
\end{equation}
which captures the essential behavior in the UHE regime.

In Fig.~\ref{ScalW} we show this $W_{if}$ for the pion when $\chi_p = 0.01$ (a), and  $\chi_p = 0.1$ (b).  
We also show $W_{if}$  for $\rho$-mesons when $\chi_p = 0.1$ (c) and  $\chi_p = 1$ (d)
with various Landau numbers $n_i$.
We see that $W_{if}$ is not changed for the large numbers of Landau levels, $n_i \gtrsim 10^5$.
As discusses later, the filled circles indicate the energy at the adiabatic limit \cite{P2PiHe}.
\begin{figure}[ht]
\vspace{0.5cm}
\begin{center}
{\includegraphics[angle=270, scale=0.55]{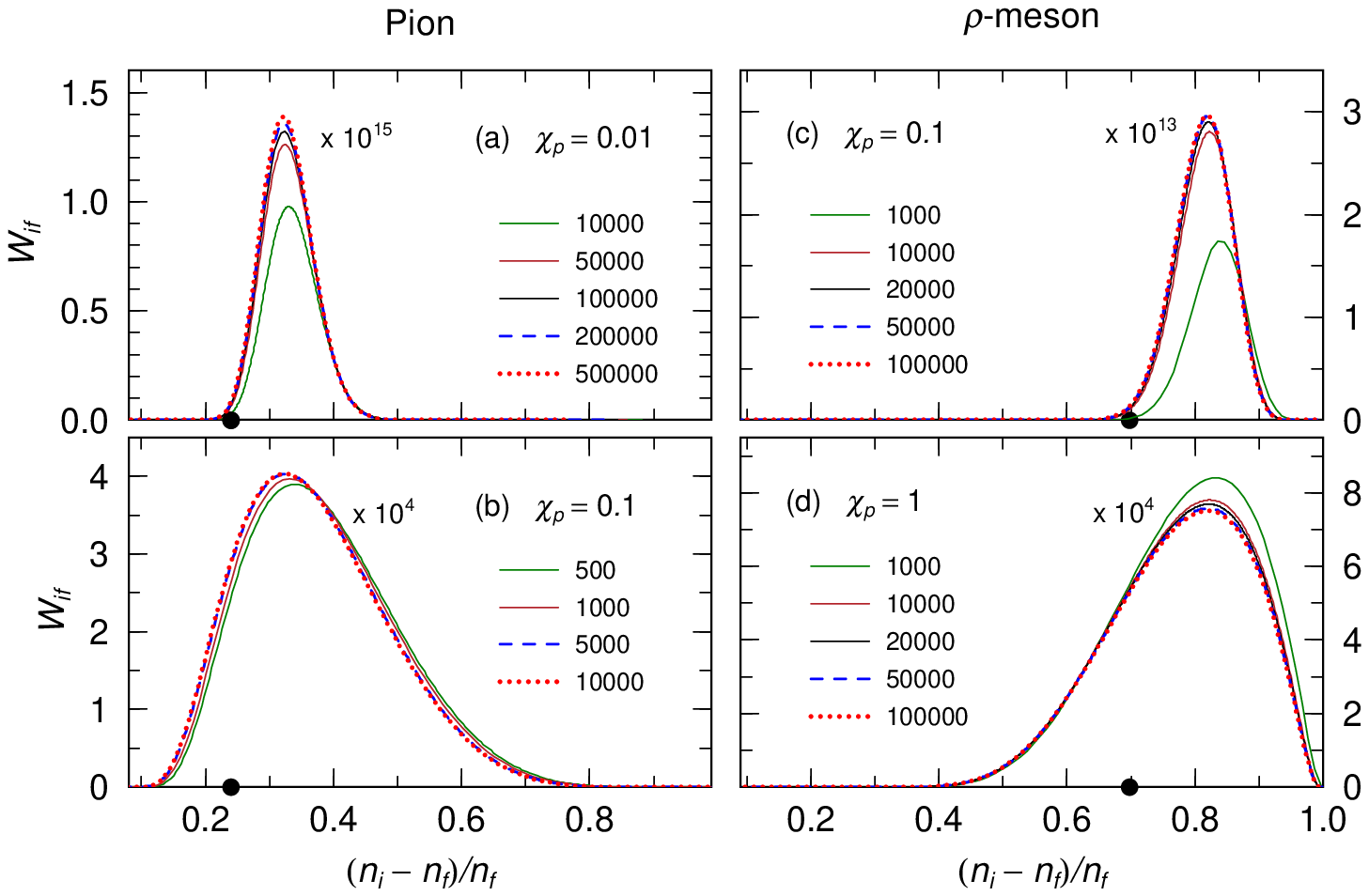}}
\caption{\small
The scaling  relation of $\cW_{if}$ in Eq.~(\ref{Wif})
when $\chi_p=0.01$ (a), and $\chi_p=0.1$ (b) for $\pi^0$, 
 $\chi_p=0.1$ (c), and  $\chi_p=1$ (d) for $\rho^0$.
Black filled circles indicate the energy of the adiabatic limit.
The numbers shown in the key correspond 
to the Landau levels represented by each line.}
\label{ScalW}
\end{center}
\end{figure}

As demonstrated in Ref.~\cite{P2PiHe}, when the strength is concentrated in the momentum region, 
the final proton and emitted particles have momenta aligned 
with the incident proton momentum in the UHE region.
In this case, the $\cT_{if}$ can be approximately expressed  as
$\cT_{if} \propto \delta \left( q_z \right)$.
In the UHE region, the quantity $\cM$ can then be expressed as
\begin{equation}
\cM^2 \approx 2 \pi \sqrt{ \frac{eB} { n_i } }  \cW_{if} (\chi_p) \delta \left( q_z \right) .
\end{equation}

From this we obtain
\begin{eqnarray}
\frac{d^3 \Gamma_{A}}{d q^3} &=&  
\frac{d_A g_A^2 \sqrt{eB} }{8 \pi e_q}  \sum_{n_f} \frac{ \cA_{if} }{E_{iT} E_{fT}} 
\frac{ \cW_{if} (\chi_p) }{ \sqrt{n_i} } 
 \delta(E_{fT} + e_q - E_{iT}) \delta(q_z ) .
\label{WidHE1}
\end{eqnarray}
%
%
Then, the total decay width is given by
\begin{eqnarray}
\Gamma_{A} &=&  
\frac{d_A }{4} g_A^2 \sqrt{eB}  \sum_{n_f} \frac{ \cA_{if} }{E_{iT} E_{fT}} 
\frac{ \cW_{if} (\chi_p) }{ \sqrt{n_i} } .
\label{TotWdHE}
\end{eqnarray}
Since the above equations are derived in  the frame where $p_{iz} = 0$, 
we write $E_{i(f)T} = \sqrt{ 2 eB n_{i(f)} + M_p^2}$ for the energy $E_{i(f)}$.

On the other hand, for large  $n_i$ and $n_f$, we can treat $n_f$ as a continuous variable and 
convert the summation over $n_f$ to an integral as
$eB \sum_{n_f} \equiv \int E_{fT} d E_{fT}$.
Then, we obtain
\begin{eqnarray}
\frac{d \Gamma_{A}}{ d e_q} &=&  
\frac{d_A g_A^2 }{ 4 \sqrt{eB} } \frac{ \cA_{if} }{ E_i }  \frac{ \cW_{if} (\chi_p) }{ \sqrt{n_i} } .
\label{dWDdE}
\end{eqnarray}

Furthermore, using the relation $\Gamma_A (p_{iz} ) = \Gamma_A(p_{iz}=0) E_{iT} / E_i$.
we can express the differential decay width for a general $p_{iz}$ as  
\begin{eqnarray}
\frac{d^3 \Gamma_{A}}{d q^3} &=&  
\frac{d_A g_A^2 }{8 \pi} \frac{E_{iT}}{E_i} \int  d E_{fT} 
\frac{ E_{fT}  A_{if} }{\sqrt{eB} E_{iT} E_{fT} e_q} 
\frac{ \cW_{if} (\chi_p) }{ \sqrt{n_i} } \delta \left( q_z - \frac{e_q}{E_i} p_{iz} \right) \delta \left( E_f + e_q - E_i  \right) 
\nonumber \\ &=&
\frac{d_A g_A^2 }{8 \pi } \frac{ A_{if}  }{ \sqrt{eB} E_i e_q } 
\frac{ \cW_{if} (\chi_p) }{ \sqrt{n_i} }  \delta \left( q_z - \frac{e_q}{E_i} p_{iz} \right) .
\label{WidHE2}
\end{eqnarray}
Here,  it is important to note that both $e_q \frac{d}{dq^3}$ and $\delta (e_q - \cdots) \delta(q_z  - \cdots)$ remain 
covariant under the Lorentz transformation along the $z$-direction.
By performing the integration over the polar angle, we finally get
\begin{eqnarray}
\frac{d \Gamma_{A}}{ d e_q} &=&  
\frac{d_A g_A^2 }{ 4 \sqrt{eB} }  \frac{ A_{if} }{ E_i }  \frac{ \cW_{if} (\chi_p) }{ \sqrt{n_i} } .
\label{dWdEpz}
\end{eqnarray}


Here, we discuss the decay width in the UHE limit $M_q /E_i \rightarrow 0$ at $p_{iz}=p_{fz}=q_z=0$.

When $E_i \approx \sqrt{2 eB n_i}  \gg M_q$, $\sqrt{2en_i B} = \sqrt{E_i^2 - M_p^2} \approx E_i - M_p^2/2 E_i$, and
\begin{equation}
\cA_{if} =  E_i E_f - 2 eB \sqrt{n_i n_f} - M_p^2 
\approx \frac{M_p^2 E_f }{2E_i} \frac{M_q^2 E_i }{2E_f} - M_p^2 = \frac{M_p^2 (E_i - E_f)^2}{2 E_i E_f} .
\label{AifUR}
\end{equation}  
Then,
\begin{eqnarray}
\frac{d \Gamma_{A}}{ d e_q} &\approx&  
\frac{d_A g_A^2 }{ 2 \sqrt{2} }  \frac{ M_p^2 e_q^2 }{ E_i^3 E_f }  \cW_{if} (\chi_p)  .
\label{dWdEpUR}
\end{eqnarray}

\newpage

\section{Results}
In this section, we present numerical results for $\gamma$, $\pi^0$ and $\rho^0$ synchrotron emission 
from protons with energies ranging from $1$ to $10^{9}$~TeV 
in magnetic fields of $B = 10^{13}, 10^{14}$, and $10^{15}$~G. 
The maximum Landau level in these conditions reaches approximately $10^{29}$.
We set the $\rho N$ coupling to be the same as that of $\pi N$ coupling, 
i.e. $g_{\rho} = g_{\pi}$.

\begin{figure}[ht]
\begin{center}
{\includegraphics[scale=0.55]{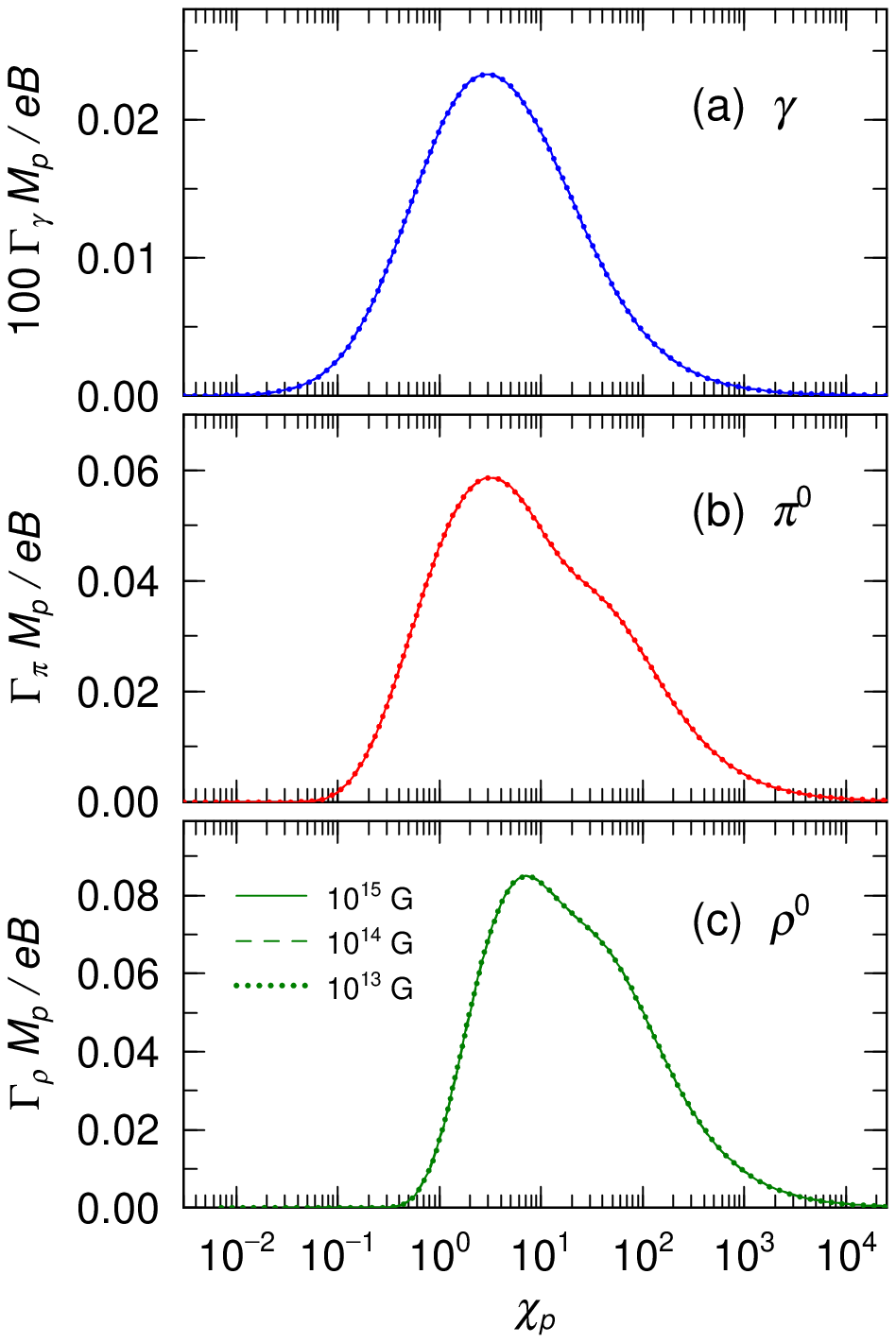}}
\caption{\small
Total decay width of  a proton at $p_{iz} = 0$ for the direct $\gamma$ (a), $\pi^0$- (b) 
and $\rho^0$- meson
emission versus $\chi_p$.
The solid, dashed and dotted lines represent the results at $B=10^{15}$, $10^{14}$ and $10^{13}$~G, respectively. 
}
\label{TWid}
\end{center}
\end{figure}

\subsection{Proton Decay Width}
Figure~\ref{TWid} shows the total decay widths of protons for direct $\gamma$, $\pi^0$-, and $\rho^0$-meson  emission as functions of $\chi_p$. 
The decay widths are scaled by a factor of $eB/M_p$, leading to identical values across different magnetic field strengths. 
These results demonstrate that the total decay widths, when divided by the magnetic field strength, depend only on $\chi_p$ 
and are thus proportional to $B$.

\begin{figure}[ht]
\centering
{\includegraphics[angle=270, scale=0.45]{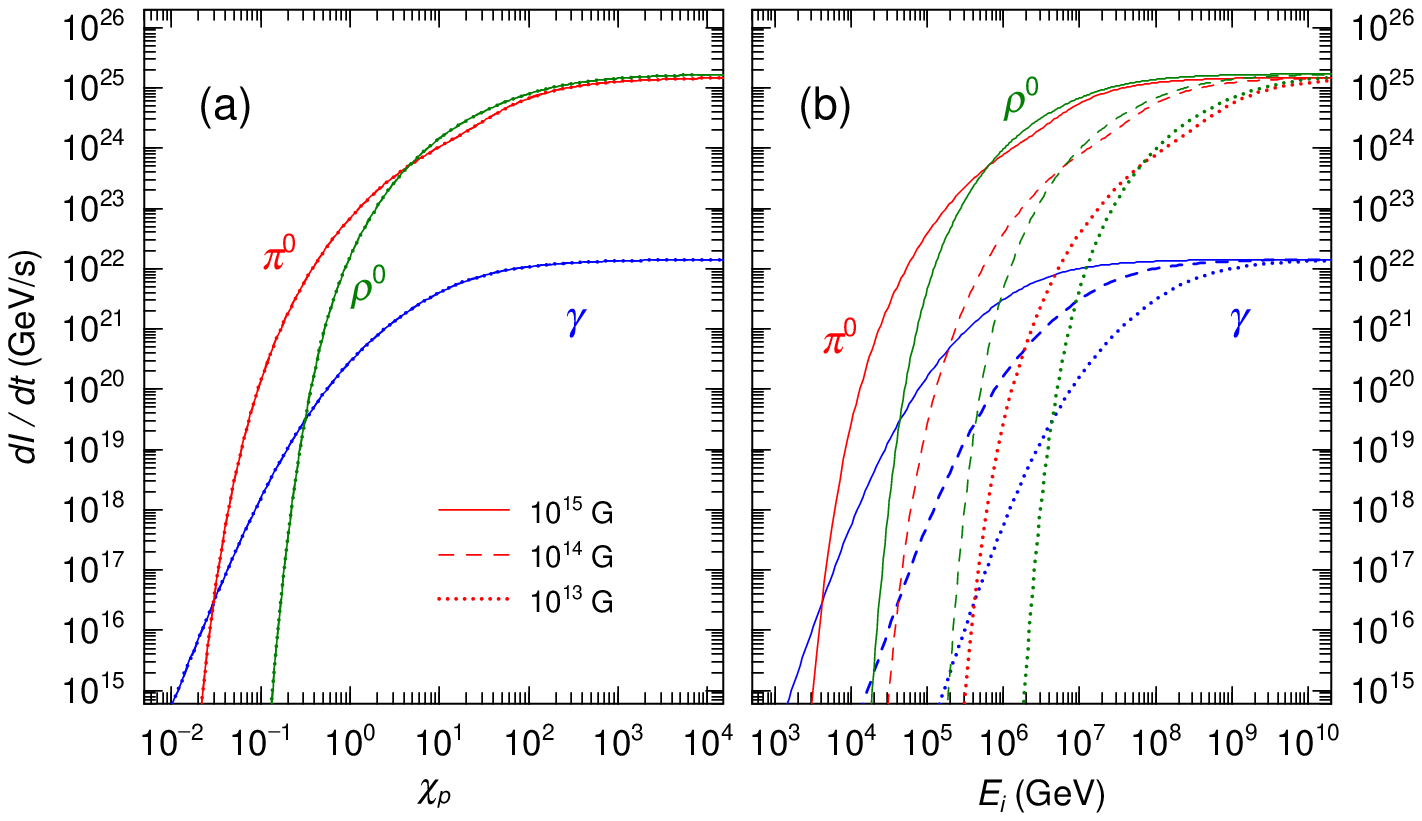}}
\caption{\small
Total synchrotron luminosity of  the direct $\gamma$, $\pi^0$- and $\rho^0$-mesons
versus $\chi_p$ (a) amd $E_i$ (b)
The red, blue and green lines represent the results for $\gamma$, $\pi$ and $\rho$-mesons,
and the solid, dashed and dotted lines indicate those at $B=10^{15}$, $10^{14}$ and $10^{13}$~G, respectively. 
}
\label{TLum}
\end{figure}

From this figure, we see that the peak positions of $\chi_p$ are
about $\chi_p \approx 1 - 10$ ($E_i = 0.14 -1.5$~PeV at $B=10^{15}$~G), and  they are
not significantly different from that of $\gamma$, $\pi^0$, and $\rho^0$ emission. 
The peak height of the $\gamma$ decay width is approximately two orders of magnitude smaller than those of $\pi^0$ and $\rho^0$, 
indicating that the overall magnitude is primarily determined by the coupling strength. 
As a result, the functional form of the total decay width is similar for all three channels.

In  Fig.~\ref{TLum}, we show the total luminosities for direct $\gamma$, $\pi^0-$, and $\rho^0$-meson emission,  
which are defined as
\begin{equation}
\frac{dI}{dt} = \int d e_q \, e_q \frac{d\Gamma}{d e_q} ,
\end{equation}
at $B= 10^{13}$~G, $B= 10^{14}$~G and $B= 10^{15}$~G.
In the left  (a) and right (b) panels  the results are plotted as   
functions of $\chi_p$ (a) and the  incident proton energy $E_i$ (b), respectively. 
We see  that the luminosity  is independent of the magnetic field strength at fixed $\chi_p$. 
This indicates that the luminosities are universal functions of $\chi_p$, regardless of the value of $B$.

%
Next, we  present the differential decay width for pionic emission from protons at magnetic field strengths of $B = 10^{13}, 10^{14}$, and $10^{15}$~G.
As shown in Fig.~\ref{dWdE-Pi}, all the results coincide when divided by $B^2$, demonstrating 
that the differential decay width is proportional to $B^2$ at fixed $e_q / E_i$.

\begin{figure}[htb]
\centering
{\includegraphics[scale=0.4, angle=270]{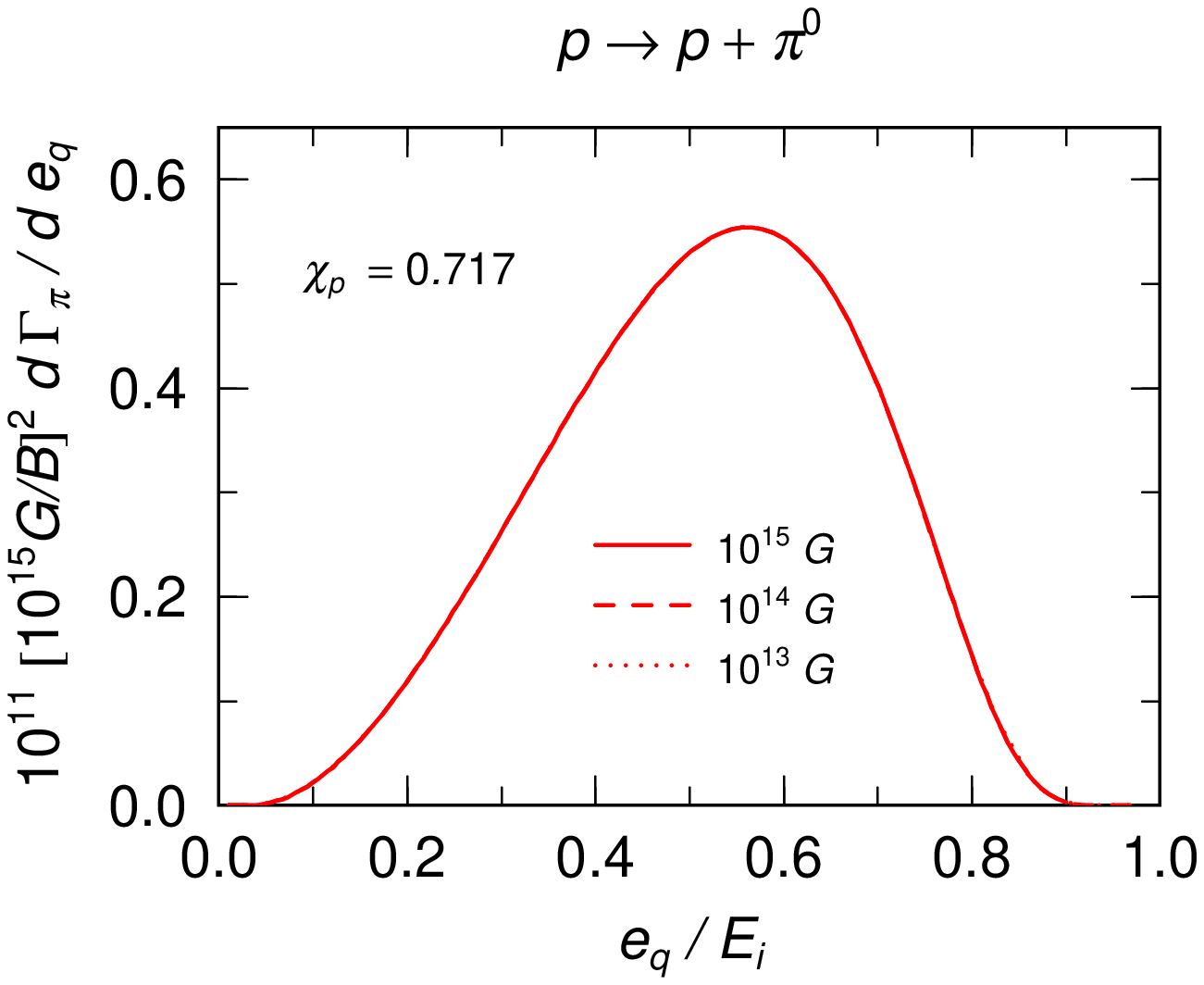}}
\caption{\small
Differential decay width $d \Gamma / d e_q$  (multiplied by $[(10^{15}~{\rm G})/B]^2$) for
proton to pion decay at $p_{iz} = 0$  
versus the emitted pion energy $e_q$ normalized by the incident proton energies $E_i$ for $\chi_p =0.717$.
}
\label{dWdE-Pi}
\end{figure}
%
 
In Fig.~\ref{dWdE}, we show the differential decay widths at $B = 10^{15}$~G for direct $\gamma$ (a), 
$\pi^0$ (b), and $\rho^0$-meson emission (c). 
The results are displayed for several incident proton energies: 
$E_i = 1$~TeV, 10~TeV, 100~TeV, 1~PeV, and 10~PeV. 
These plots illustrate how the energy spectrum of the emitted particles evolves with increasing $E_i$.
There is no strength for the $\rho^0$-emission at $E_i = 1$~TeV, 
which is lower than the energy threshold.

\begin{figure}[htb]
\begin{center}
{\includegraphics[scale=0.5]{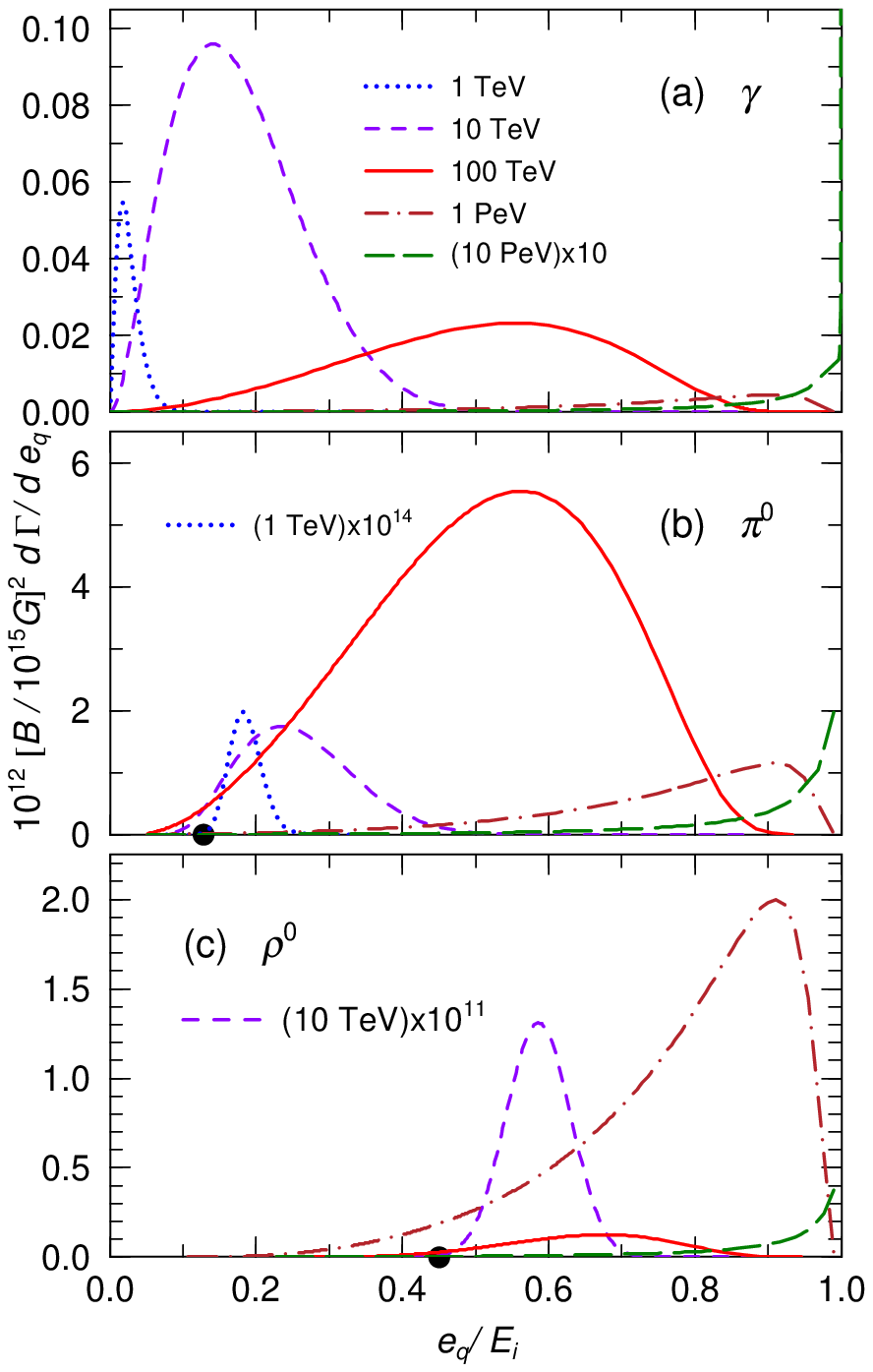}}
\caption{\small
Differential decay width $d \Gamma / d e_q$ of protons for direct $\gamma$ (a), $\pi^0$ (b), 
and $\rho^0$ emission (c) at $B=10^{15}$~G.. 
The blue dotted, purple dashed, red solid, brown dot-dashed and green long-dashed lines 
represent the results for $E_i = 1$~TeV, 10~TeV, 100~TeV, 1~PeV, and 10~PeV. 
}
\label{dWdE}
\end{center}
\end{figure}

In addition, the solid filled circles indicates the energy at the adiabatic limit \cite{P2PiHe},
which is also presented in Fig.~\ref{ScalW}.
In the adiabatic limit, it can be assumed that the relative momentum
between the final proton and the pion is zero, and that
the two particles move with nearly the same velocity.
In this case, the ratio of their energies
is approximately given by the ratio of their masses: $e_a / E_f \approx m_a / M_p$,
and the final proton and emitted particle energies are given by
\begin{equation}
E_f \approx \frac{M_p}{M_p + m_a} E_i , ~~
e_a \approx \frac{m_q}{M_p + m_a} E_i .
\end{equation}

When the incident proton energy exceeds the threshold, particles are emitted at energies near the adiabatic limit.
As the mass of the emitted particle becomes larger, the threshold energy also increases.  
However, the threshold energy cannot be derived analytically in the present approach, and in the numerical calculations it is found to be much larger than $m_a + M_p$.
As the incident proton energy increases, 
the differential decay width spreads over a broader energy region (cf. Figure \ref{dWdE}).  
The peak energy of the distribution rises, 
and the minimum energy shifts slightly below the adiabatic-limit energy.
When the proton energy increases further, reaching around 10~TeV, 
the emitted particles carry away most of the energy and 
their energies become nearly equal to that of the incident proton.
In this regime, the energy distribution becomes narrow, and the decay width becomes very small.


\subsection{Luminosity Distribution of Emitted Particles}

The luminosity distribution of the emitted particles,$d(d^3 I / d q^3)/dt = e_q d^3 \Gamma_{p \pi} / d q^3$ 
is obtained by averaging over the incident proton angle, 
where the proton momentum distribution has a spherical symmetry.
The angle averaged luminosity distribution becomes
\begin{eqnarray}
\left< \frac{d \Gamma_A}{d q^3} \right> &=& 
\frac{\int dp_{iz} \sum_{n_i} \delta \left( E_i - \sqrt{p_{iz}^2 + 2n_i  + M_p^2} \right)
\frac{d_A g_A^2 }{8 \pi } \frac{ A_{if}  }{ \sqrt{eB} E_i e_q } 
\frac{ \cW_{if} (\chi_p) }{ \sqrt{n_i} }  \delta \left( q_z - \frac{e_q}{E_i} p_{iz} \right) 
}{ \int dp_{iz} \sum_{n_i} \delta \left( E_i - \sqrt{p_{iz}^2 + 2n_i + M_p^2 } \right) }
\nonumber  \\ &=&
 \frac{d_A g_A^2 }{16 \pi } \frac{ A_{if} }{ e_q^2 \sqrt{ E_i^2 - M_p^2 } } 
\frac{ \cW_{if} (\chi_p) }{ \sqrt{eB n_i} }  ~~.
\end{eqnarray}

\begin{figure}[bht]
\centering
{\includegraphics[angle=270,  scale=0.45]{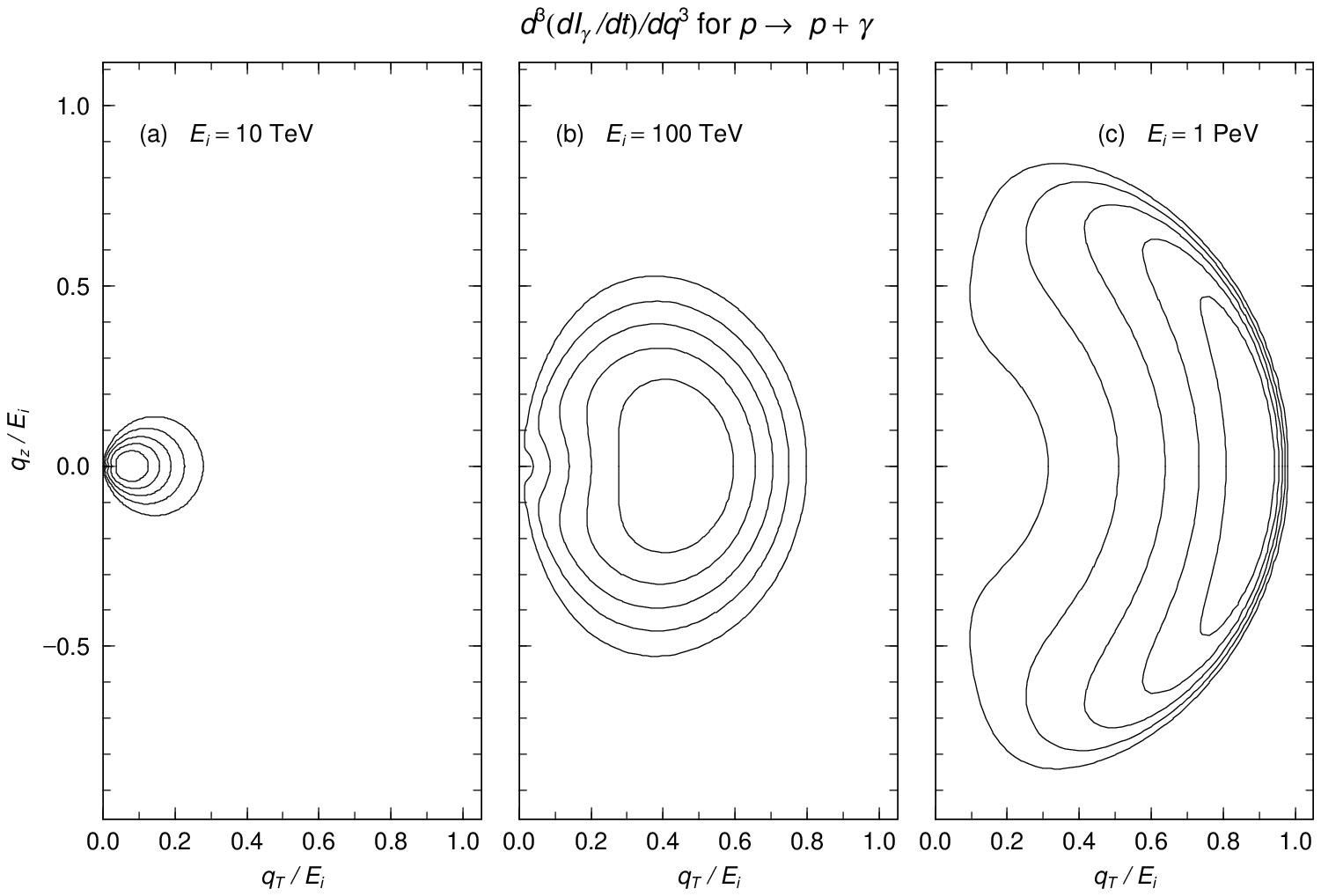}}
\caption{\small
Contour plot of the differential photon luminosity for 
$E_i = 10$~TeV (a), 100~TeV (b) and 1~PeV (c)  at $B=10^{15}$~G.
They are integrated over the incident proton angle.
Lines on each plot show the contours of relative strength, 
17\%, 33\%, 50\%  67\% and 83\%.
The vertical and  horizontal axes are the $z$-component and the
transverse component of the emitted photon momentum.
 }
\label{PhPd}
\end{figure}
\begin{figure}[htb]
\centering
{\includegraphics[angle=270,  scale=0.45]{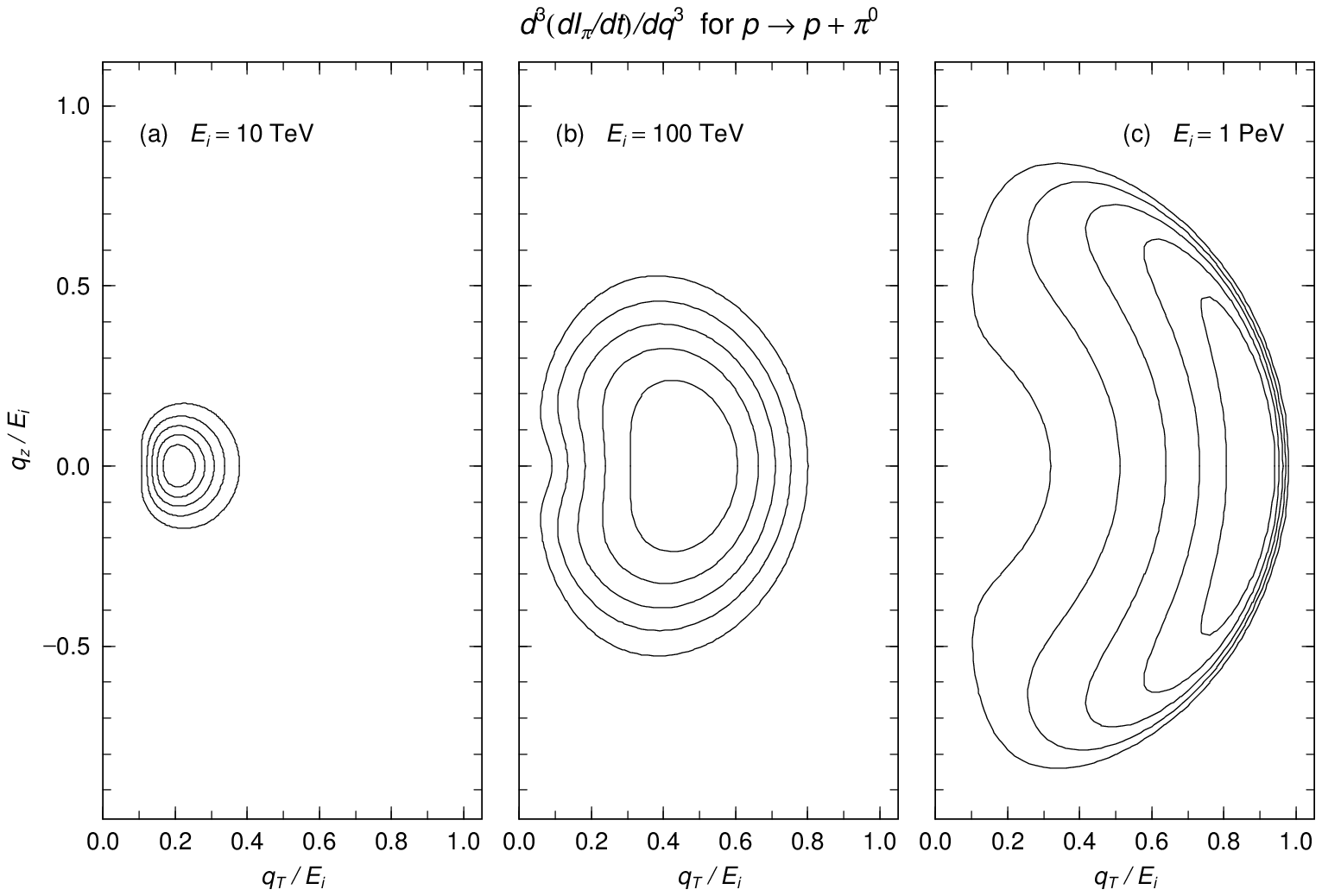}}
\caption{\small
Contour plot of the differential $\pi^0$ luminosity for $E_i = 10$~TeV (a), 100~TeV (b) 
and 1~PeV (c).
Same as Fig.~\ref{PhPd}
 }
\label{PiPd}
\end{figure}
\begin{figure}[bht]
\centering
{\includegraphics[angle=270,scale=0.45]{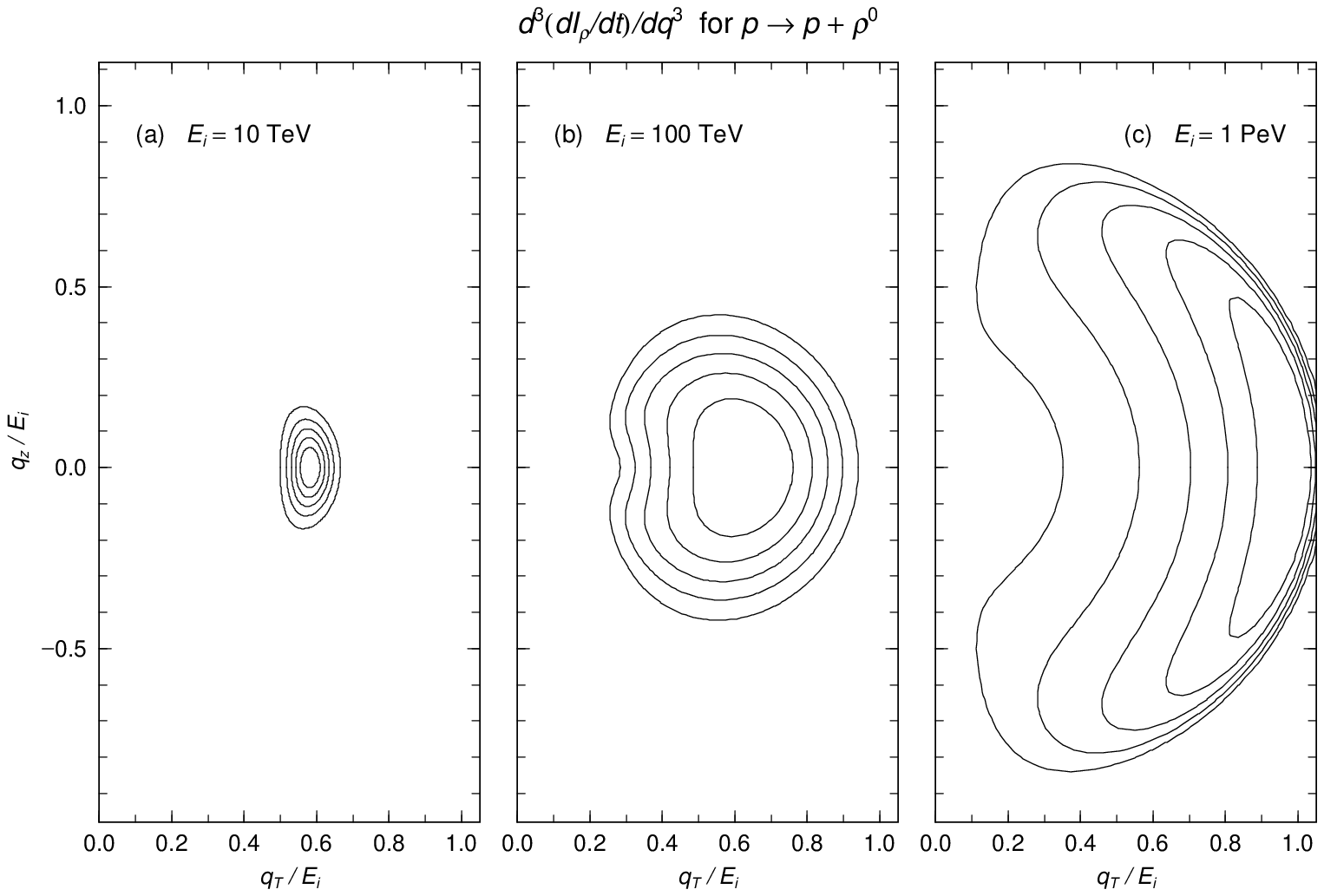}}
\caption{\small
Contour plot of the differential $\rho^0$ luminosity for $E_i = 10$~TeV (a) , 100~TeV (b) and 1~PeV (c).
Same as Fig.~\ref{PhPd}
 }
\label{RhPd}
\end{figure}


In Figs.~\ref{PhPd}, \ref{PiPd} and \ref{RhPd}, we present contour plots 
of the luminosity distributions for photons, $\pi^0$s and $\rho^0$s, respectively,
for incident energies $E_i = 10$~TeV (a), 100~TeV (b), and 1~PeV (c)  at $B=10^{15}$~G. 
All results are similar.
At $E_i=10$~TeV, the momentum of the emitted particles is only distributed over a very narrow range 
around the energy the adiabatic limit.
However, as the energy increases,  the momenta of the emitted particles is distributed over a larger range.

The drift angles of the emitted particles closely match those of the incident protons, and the energy distribution becomes narrower as the emission angle  increases.
Very little emission occurs along the direction of the magnetic field.
Moreover, emission at lower energies decreases as the incident proton energy increases, consistent with the results shown in Fig.\ref{dWdE}.


\subsection{Comparison with Semiclassical Theory}

Within a semiclassical framework, Berezinsky et al.~\cite{BDK95} derived the total decay width as
\begin{equation}
\Gamma_{\pi} = \frac{3 g_{\pi}^2 }{2^6 \pi} \frac{M_p^2}{E_i} \chi_p^{2/3}
= \frac{3 g_{\pi}^2 }{64 \pi} \frac{eB}{M_p} \chi_p^{-1/3}.
\end{equation}
Here, the total decay width is proportional to the magnetic field strength, which is consistent with our result.
For $B = 10^{15}$~G, this expression gives 
$\Gamma_{\pi} = 1.77 \times 10^4 \chi_p^{-1/3}$~(eV), 
which is considerably larger than the values obtained in our calculation.

It should be noted, however, that their derivation assumes the energy of the emitted pions 
to be negligibly small, i.e., $e_q \ll E_i$.
In contrast, our results show that the energies of the emitted particles 
are distributed over a broad range and can be comparable to the incident proton energy.
In the high-energy limit, the emitted particle energy approaches that of the initial proton, 
so the particle luminosity does not increase substantially.

\begin{wrapfigure}{r}{8cm}
\centering
\vspace{-0.5cm}
{\includegraphics[scale=0.42,angle=270]{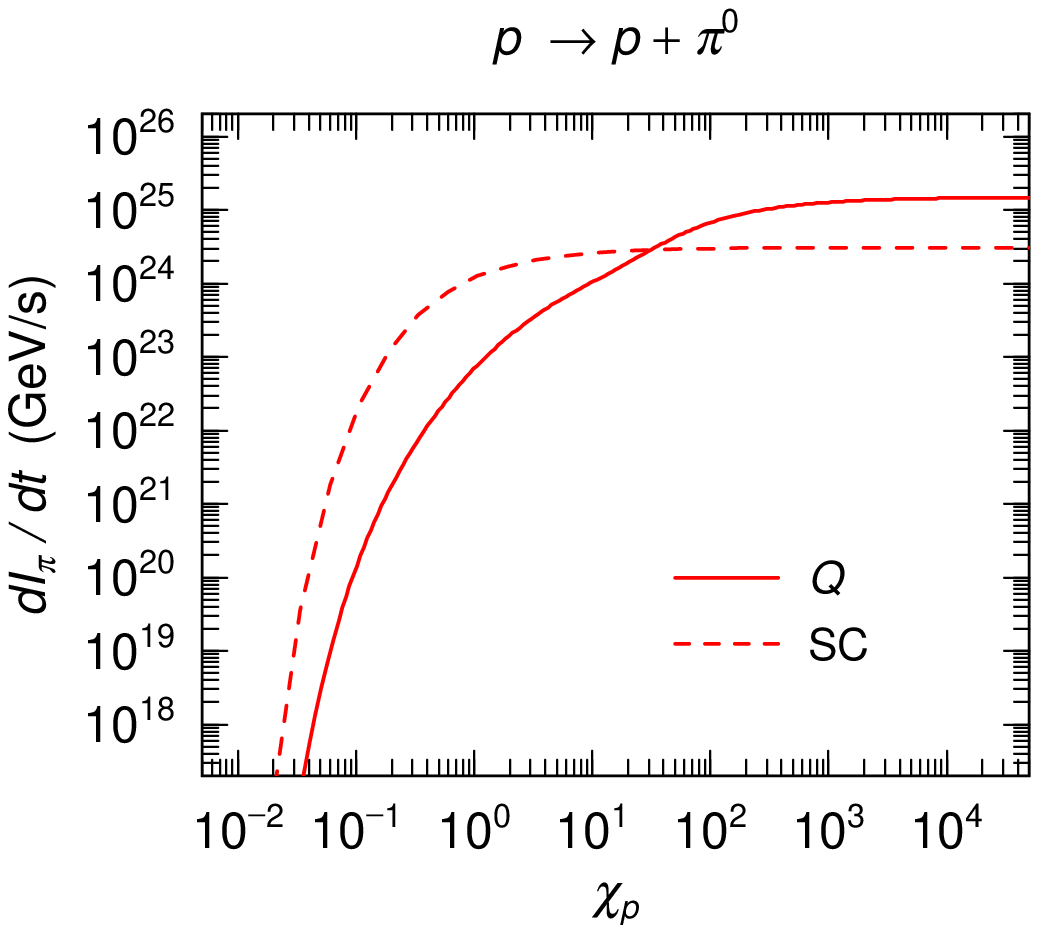}}
\caption{\small
Total pion luminisity as a function of $\chi_p$.
The solid and dashed lines represent the relativistic quantum 
and the semiclassical results in Ref.~\cite{TK99}.
 }
\label{TotLumSC}
\end{wrapfigure}

Another semiclassical approach was presented in Ref.~\cite{TK99}, 
where the total luminosity and energy distribution of the emitted particles were discussed. 
The luminosities of the emitted particles can be expressed as a function of $\chi_p$, 
and their magnitudes are independent of the magnetic field strength. 
These features are consistent with our results.

In Fig.~\ref{TotLumSC},
we compare the total luminosities of pion emission as a function of $\chi_p$ 
in our relativistic quantum approach (solid line) 
with those from the semiclassical approach of Ref.~\cite{TK99} (dashed line). 
In both approaches, the luminosities are independent of $B$. 
At small $\chi_p$ (low-energy region), the semiclassical results are larger 
and rise more steeply than those of the quantum theory. 
At high energies, however, the two approaches converge to finite values, 
with the semiclassical prediction becoming slightly smaller than the quantum one.

While the total luminosities show similar trends, 
differences appear in the detailed energy distributions of the emitted pions. 
Fig.~\ref{dUdEpiC} shows the distributions at $B=10^{15}$~G 
for $E_i = 100$~TeV, 10~PeV, 1~EeV, and 10~EeV, comparing our quantum model with Ref.~\cite{TK99}. 
In the semiclassical theory, the low-energy part of the spectrum is more pronounced 
and tends to broaden toward lower energies as $E_i$ increases, 
whereas in the relativistic quantum calculation the low-energy component is suppressed.

\begin{wrapfigure}{r}{8cm}
\centering
\vspace{-0.5cm}
{\includegraphics[scale=0.4]{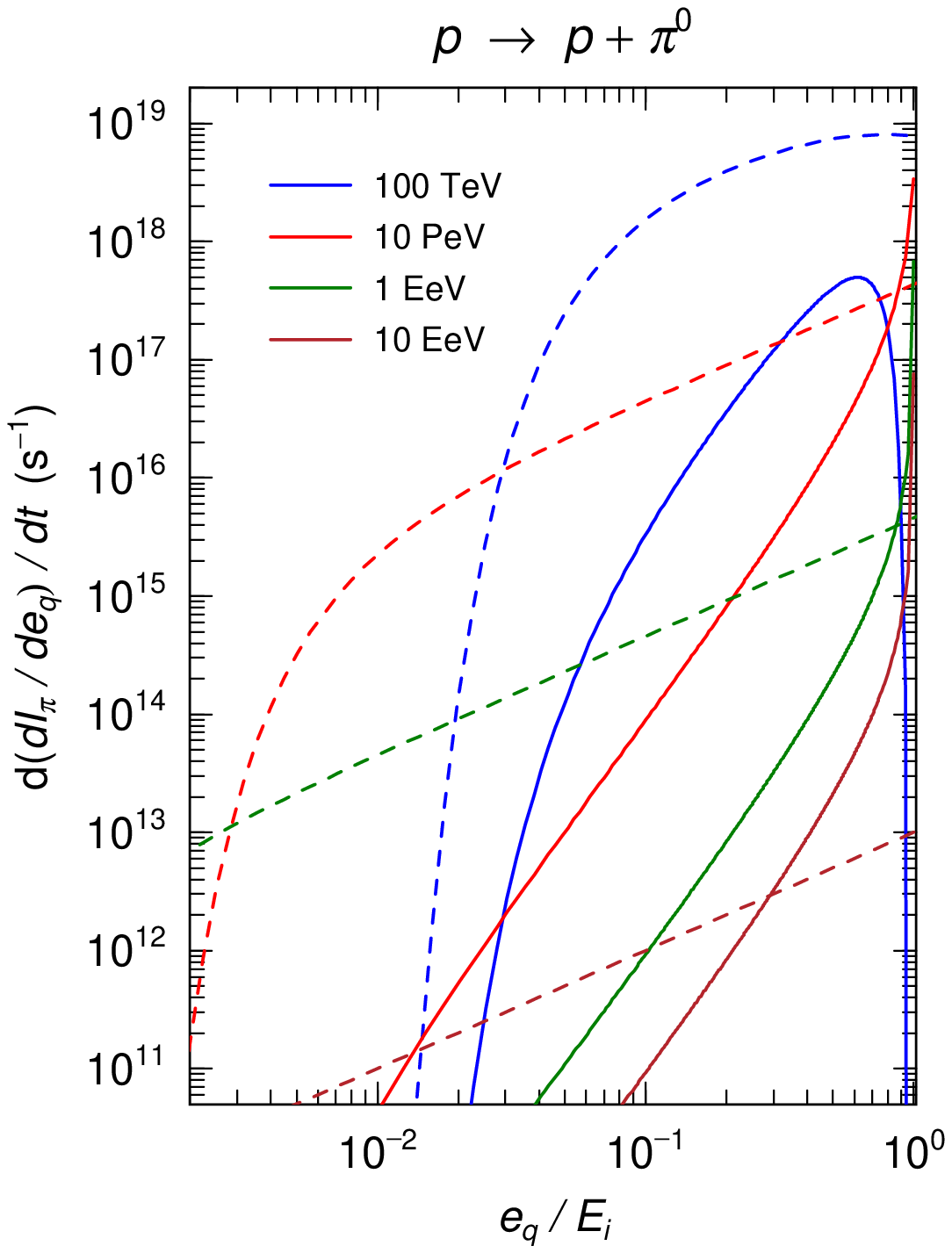}}
\caption{\small
Diffrential  pion luminisity with respect to the emitted pion energy at $B=10^{15}$~G.
The blue, red, green and brown lines represent results with $E_i =  100$~TeV,
10~PeV, 1~EeV and 10~EeV, respectively.
The solid and dashed lines indicate  our results  
and the semiclassical results in Ref.~\cite{TK99}, respectively.
 }
\label{dUdEpiC}
\end{wrapfigure}
	
In the semiclassical theory, the low-energy part of the pion spectrum is more pronounced than in the quantum calculation. 
As the incident proton energy increases, the semiclassical distribution broadens toward the low-energy side, 
whereas in the quantum case the low-energy component is suppressed. 	
For $E_i = 100$~TeV, the pion luminosity is higher in the semiclassical theory than in the quantum theory.
When the pion energy approaches the incident proton energy, 
the quantum calculation predicts that the luminosity vanishes at $e_q = E_i$, 
while the semiclassical theory continues to give a finite value.

In the quantum calculation, recoil effects suppress the spectrum near the kinematic endpoint, 
so the luminosity should vanish as $e_q \to E_i$. 
In our practical computation, however, we rely on a scaling rule to handle the very large Landau numbers. 
For incident energies $E_i \gtrsim 1$~PeV, the actual Landau number exceeds $10^{18}$, 
but in the numerical evaluation we calculate up to $n_{i,f} \sim 10^{5}$ and then use the scaling rule 
to extrapolate to the physical regime. 
In this limit, the Landau-level spacing $\Delta n_{if}$ becomes extremely small, 
and the scaling approximation cannot faithfully reproduce the endpoint behavior. 
As a result, the calculated pion luminosity does not decrease to zero in the region $e_q \lesssim E_i$, 
contrary to the exact quantum expectation. 
This discrepancy should therefore be regarded as a limitation of the scaling-rule approximation 
rather than a genuine physical effect.


Furthermore, the semiclassical calculation exhibits a different trend for gamma rays and $\rho$ mesons.  
In this approach, their luminosities increase continuously with the proton energy, 
showing no evidence of saturation.  
This behavior indicates a significant difference between the quantum and semiclassical approaches for vector particle production.  
In contrast, the discussion of pion emission demonstrates that the semiclassical theory fails even to approximate the results of the quantum calculation.  

%


\section{Summary}

In this work, we investigated synchrotron radiation from ultra-high-energy protons propagating 
in strong magnetic fields within a fully relativistic quantum framework.
We solved the Dirac equation in the presence of a strong magnetic field and 
derived the proton propagator from its solution.
While evaluating synchrotron radiation, we found a scaling rule for the overlap integral 
of two harmonic oscillator wave functions in Eq.~(\ref{Wif}).
Using this scaling property, we derived the decay width of a proton occupying very large Landau levels in a fully relativistic and quantum mechanical manner.
Furthermore, we found that the proton decay width adheres to a straightforward scaling rule, 
whereby its ratio to the magnetic field strength is solely dependent 
on the product of the incident proton energy and the magnetic field strength.

In the same classical theory, the radiation probability increases monotonically with the incident energy~\cite{Jackson}.  
In contrast, our quantum calculation shows that the decay width peaks and then rapidly decreases, and the particle luminosity saturates to a finite value for proton energies exceeding approximately 10~PeV at \( B = 10^{15}~\mathrm{G} \).  
In this ultra-high-energy region, the discrete nature of Landau levels becomes negligible, indicating that our quantum approach approaches the classical limit.  
However, classical theory does not properly account for the recoil of the emitting proton and assumes that the energy of the emitted particle is much smaller than the incident proton energy.  
In our quantum framework, the emitted particle energy is distributed around the adiabatic limit, and the peak of this distribution shifts to larger energy as the incident proton energy increases.  
When the proton energy exceeds about 10~PeV, the emitted energy is concentrated near the incident energy, indicating that recoil effects become significant in this regime.  
These effects are naturally incorporated in the quantum theory.  
This is because classical theory cannot treat the emitted radiation as a particle and therefore cannot fully describe the recoil effect.  
In contrast, our approach is based on quantum theory, where the emitted radiation is treated as a real particle, and the recoil is handled exactly through energy-momentum conservation.

Semiclassical approaches to synchrotron radiation, such as those developed by Sokolov and Ternov~\cite{SokolovTernov} or Erber~\cite{Erber1966}, incorporate quantum features like Landau quantization while treating the radiation field classically. 
These methods can approximate certain quantum effects, but do not fully capture essential features such as recoil and photon quantization, which are naturally included in our fully relativistic quantum treatment. 
Semiclassical theories have also been extended to describe meson radiation, 
such as for $\pi$ and $\rho$ mesons, 
as modified analogues of the theory of optical emission. 
However, as demonstrated in this work, when the emitted particles are massive, the recoil becomes significant, and current semiclassical frameworks apparently fail to accurately describe the synchrotron radiation process.


As shown in our numerical analysis, both the luminosity and 
the decay width are governed by the single scaling variable
$\chi_p = e B E_i/M_p^3$.
For fixed $\chi_p$, the luminosities become independent of $B$, whereas the
decay widths scale linearly increases with it.  
When the value of $\chi_p$ becomes very large
($\chi_p \gtrsim 10^2$), the luminosities approach constant values, while
the decay width decreases towards zero as $\chi_p$ approaches infinity.
In this regime, emitted particles are produced only rarely, but when they
are produced their energy becomes comparable to that of the incident
proton.  
This behaviour occurs when both the incident proton energy and the magnetic field strength are increased.

However, when only the magnetic field is increased while keeping the proton
energy fixed, the Landau number of the incident proton decreases and
quantum effects become significant.  
In the limit of an infinitely large magnetic field, the proton is forced into the lowest Landau level, and the available phase space for the emitted particle shrinks rapidly.  
Consequently, both the luminosity and the decay width must decrease 
in the limit of extremely
large magnetic fields, even though the scaling variable $\chi_p$ diverges.

We also note that in heavy-ion collisions, peripheral ultra-relativistic
collisions at the Relativistic Heavy Ion Collider (RHIC) and Large Hadron Collider (LHC) are expected to generate magnetic fields as large as
$10^{18} –10^{19}$~G \cite{SIT2009,RHIC+LHC25}.  
For example, at $E_i = 100$~GeV and $B = 10^{19}$~G, we obtain
$\chi_p \simeq 7.1$ and $n_i \simeq 290$ for a proton moving
perpendicularly to the magnetic field.  
In our approach the corresponding decay widths are
estimated to be $\Gamma_\gamma \approx 2.7 \times 10^{-5}$~GeV,
$\Gamma_\pi \approx 4.8 \times 10^{-3}$~GeV,
and $\Gamma_\rho \approx 1.1 \times 10^{-2}$~GeV.
Furthermore, as the Landau number is not large, the scaling behavior derived in the high-energy limit will not be realized.
The actual decay widths must therefore be smaller than the above values. 
We therefore expect that the effect of synchrotron radiation on the observed quantities in ultra-relativistic heavy-ion collisions is minimal.

Our approach enables the calculations of synchrotron radiation 
in the high-energy region above several hundred GeV, 
including the full momentum distribution of the emitted particles.
In future work, we plan to compute the momentum distributions of gamma rays emitted from ultra-high-energy cosmic rays and apply these results to analyze their astrophysical sources.
This will involve integration over the momentum distribution of the incident protons, which is not spherically symmetric \cite{Allardt2022, Allardt2024}.

In our approach, the radiation probability vanishes in the direction parallel to the magnetic field, in agreement with the classical prediction.  
In this study, we assume that the drift angle of the emitted particle is equal to that of the incident proton.  
This assumption is numerically supported in the case where the drift angle is zero, corresponding to emission perpendicular to the magnetic field \cite{P2PiHe}.  
Nonetheless, it is known that charged particles moving along the magnetic field can emit circularly polarized photons.  
In particular, Ref.~\cite{GamGene} demonstrated that photon vortices, described by Bessel wave structures, can be generated in this direction through transitions between different Landau levels.  
Such radiation originates purely from quantum effects, indicating the existence of a distinct high-energy limit not addressed in the present analysis.  
Although these contributions are expected to be small, they may still be observable and could provide an interesting direction for future investigations.  
Even for ultra-relativistic protons, quantum calculations remain applicable when the energy of the emitted particle is much smaller than that of the incident proton~\cite{GamG2023}.  
A more detailed study of this type of radiation is planned for future work.

\acknowledgements
This work was supported by Grants-in-Aid for Scientific Research of JSPS (24K07057 and 25K17403). 
It was also supported in part by the U.S. National Science Foundation grant (PHY-2411495),
the National Research Foundation of Korea (RS-2021-NR060129, RS-2025-16071941),
and the National Key R\&D Program of China (2022YFA1602401) and the National Natural Science Foundation of China (12335009, 12435010).
Work of GJM suported by the US Department of Energy under Nuclear Theory Grant DE-FG02-95-ER40934.


\end{document}